\documentclass[prl,twocolumn,showpacs,preprintnumbers,amsmath,amssymb]{revtex4}

\usepackage{graphicx}
\usepackage{dcolumn}
\usepackage{bm}
\usepackage{mathrsfs}
\usepackage{subfigure}
\usepackage{natbib}

\begin{document}

\title{Extended Ginzburg-Landau formalism for two-band superconductors}

\author{A. A. Shanenko}
\email{arkady.shanenko@ua.ac.be}
\affiliation{Departement Fysica, Universiteit Antwerpen,
Groenenborgerlaan 171, B-2020 Antwerpen, Belgium}

\author{M. V. Milo\v{s}evi\'{c}}
\affiliation{Departement Fysica, Universiteit Antwerpen,
Groenenborgerlaan 171, B-2020 Antwerpen, Belgium}

\author{F. M. Peeters}
\affiliation{Departement Fysica,
Universiteit Antwerpen, Groenenborgerlaan 171, B-2020 Antwerpen,
Belgium}

\author{A. V. Vagov}
\affiliation{Institut f\"{u}r Theoretische Physik III, Bayreuth
Universit\"{a}t, Bayreuth 95440, Germany}

\date{\today}

\begin{abstract}
Recent observation of unusual vortex patterns in MgB$_2$ single
crystals raised speculations about possible "type-1.5"
superconductivity in two-band materials, mixing the properties
of both type-I and type-II superconductors. However, the
strict application of the standard two-band Ginzburg-Landau
(GL) theory results in {\it simply proportional} order parameters
of the two bands - and does not support the ``type-1.5'' behavior.
Here we derive {\it the extended GL formalism} (accounting all
terms of the next order over the small $\tau= 1- T/T_c$ parameter)
for a two-band clean $s$-wave superconductor and show that the
two condensates {\it generally have different spatial scales}, with
difference disappearing only in the limit $T \to T_c$. The extended
version of the two-band GL formalism improves the validity of GL
theory below $T_c$, and suggests revisiting of the earlier
calculations based on the standard model.
\end{abstract}


\maketitle 

The Ginzburg-Landau (GL) approach~\cite{GL}, based on
Landau's theory of second-order transitions, is one of the most
powerful and most widely used theoretical tools of the present-day
physics. It constitutes a solid base for theoretical studies in
fields ranging from the condensed matter theory (e.g.
superconductivity/superfluidity, phase transitions, and fluctuation
phenomena) to particle physics and cosmology (e.g. Higgs mechanism),
and other topics reviewed in Ref.~\onlinecite{aran}. It is generally
believed that the GL theory accurately describes essential physics
in the vicinity of the critical temperature $T_c$~(and,
qualitatively, in a much wider temperature range). Surprisingly,
this is not the case for two-band (and multi-band) superconductors,
such as magnesium-diboride \cite{canf} and several iron-pnictides
\cite{pnic} etc., where the expected difference in spatial
distributions of the two Cooper-pair condensates is not captured
by the standard formulation of the two-band GL formalism. As
explained further, the latter problem requires development of
the extended GL theory, derived to a higher order in $\tau=1-
T/T_c$. This is the core objective of this Letter.

Recently, unconventional vortex patterns were observed in
single-crystal MgB$_2$ by Bitter decoration \cite{mosh} and by
Scanning SQUID microscopy \cite{nishio}. Although MgB$_2$ is largely
accepted as a type-II two-band superconductor, in Refs.
\onlinecite{mosh,nishio} no evidence of an Abrikosov lattice was found for
low vortex densities. The interpretation was offered through the
intervortex potentials derived from the standard two-band GL theory
of, e.g., Refs.~\onlinecite{zhit,gur,kosh}. Namely, for particularly
chosen (different) coherence lengths $\xi_i$ and penetration depths
$\lambda_i$ of the two bands ($i=1,2$), vortices were shown to
conventionally repel each other only at short distances, while
long-range {\it attracting}~\cite{babaev}. This gives rise
to stripe-like vortex patterns, unattainable in either type-I or
type-II superconductors, which led Moshchalkov {\it et al.}
\cite{mosh} to name this behavior ``the type-1.5
superconductivity''.

An avalanche of theoretical works followed \cite{babaev1,wang},
based either on the standard two-band GL formalism itself, or the
molecular dynamics simulations using the GL-calculated intervortex
potentials, racing to describe the new type of superconductivity.
Brandt was the first to point out that long-range vortex attraction
is not necessarily a ``type-1.5'' property~\cite{brandt}. The real
criticism followed, in the analysis of Kogan and Schmalian \cite{kog}.
They showed that in the standard formulation of the two-band GL
approach [i.e., two GL equations for two order parameters
$\Delta_i({\bf x})$ coupled through the Josephson interband coupling
terms], there appear contributions to both order parameters of
higher orders than $\tau^{1/2}$, where $\tau=1-T/T_c$. However, the
microscopic basis for the standard GL formalism assumes that only
the terms $\propto\tau^{1/2}$ are accurate, which means that
aforementioned higher-order terms are {\it incomplete and, thus,
incorrect}. After removing the higher-order contributions, Kogan and
Schmalian found the order parameters of two bands to be proportional
to each other and can be thus characterized by a single coherence
length $\xi$. As a consequence, type-1.5 superconductivity is not
supported by this formalism.

It is thus of abiding fundamental interest to clarify whether the
relation $\Delta_1({\bf x}) \propto \Delta_2 ({\bf x})$ is generic
to two-band superconductors or it holds only in the standard GL
domain (to the order $\tau^{1/2}$ in $\Delta_i$'s). To settle the
above issues, we derive here the extended version of the GL formalism
for a two-band clean $s$-wave superconductor, where the contributions to $\Delta_j$'s up to the order $\propto\tau^{3/2}$ are included in their full, correct form (while appearance of the higher orders is precluded).

Our starting point is the BCS mean-field Hamiltonian of a two-band,
$s$-wave, clean superconducting system, i.e.,
\begin{multline}
H^{\rm BCS}= H^{c} + \sum\limits_{j=1,2}\int {\rm d}^3x\,
\Bigl[{\hat \psi}^{\dagger}_{j\sigma}({\bf x})\,T_j({\bf x}) \,{\hat
\psi}_{j\sigma}({\bf x})\\+ {\hat \psi}^{\dagger}_{j \uparrow}({\bf
x}){\hat \psi}^{\dagger}_{j\downarrow}({\bf x})\, \Delta_j({\bf x})
+ {\rm h.c.}\Bigr], \label{HBCS}
\end{multline}
where $j=1,2$ denotes each of the bands, $H^c$ is the $c$-term whose
specific form (see, e.g., Ref.~\onlinecite{zhit}) is not of
relevance for the present investigation, $T_j({\bf x})$ is the
single-electron Hamiltonian, and the summation in the kinetic term
is taken over the coinciding spin indices. The generalization of the
mean-field self-consistency equation for two-band superconductors
reads
\begin{equation}
\Delta_i({\bf x})=\sum\limits_{j=1,2}\;g_{ij}\langle {\hat
\psi}_{j\uparrow}({\bf x}){\hat \psi}_{j\downarrow}({\bf x})
\rangle, \label{gap2band}
\end{equation}
with $g_{ij}$ being the relevant coupling constants
($g_{ij}=g_{ji}$). One of the most powerful formalisms to treat the
superconducting properties in the presence of a nonuniform spatial
distribution of the pair condensate are the Gor'kov equations. For
our study it is convenient to write these equations in the form of
the Dyson equation for the $2\times 2$-matrix band propagator
${\check {\cal G}}_{j \omega}$ (see, e.g., Ref.~\onlinecite{kopnin}):
\begin{equation}
{\check {\cal G}}_{j\omega}={\check {\cal G}}^{(0)}_{j\omega} +
{\check {\cal G}}^{(0)}_{j\omega}\;{\check \Delta}_j\;{\check {\cal
G}}_{j\omega}, \label{Geq}
\end{equation}
with
\begin{equation}
{\check {\cal G}}_{j\omega}= \left(
\begin{array}{cc}
{\cal G}_{j\omega} & {\cal F}_{j\omega}\\
{\bar {\cal F}}_{j\omega} & {\bar {\cal G}}_{j\omega}
\end{array}
\right), \quad {\check {\cal G}}^{(0)}_{j\omega}= \left(
\begin{array}{cc}
{\cal G}^{(0)}_{j\omega} & 0\\
0 & {\bar {\cal G}}^{(0)}_{j\omega}
\end{array}
\right), \label{Gfunc}
\end{equation}
where $\hbar\omega= \pi T(2n+1)$ is the fermionic Matsubara frequency
($n$ is an integer and $k_B$ is set to unity) and the $2\times2$ matrix
operator ${\check \Delta}_j$ in Eq.~(\ref{Geq}) is defined by
\begin{equation}
{\check \Delta}_j= \left(
\begin{array}{cc}
0 & {\hat \Delta}_j\\
{\hat \Delta}^{\ast}_j & 0
\end{array}
\right), \quad \langle {\bf x}|{\hat \Delta}_j|{\bf x}'\rangle=
\delta({\bf x}-{\bf x}') \Delta_j({\bf x}'). \label{gapcheck}
\end{equation}
Equations~(\ref{Geq}) and (\ref{Gfunc}) further give
\begin{subequations}
\begin{align}\label{Geq1}
&{\cal F}_{j\omega}={\cal G}^{(0)}_{j\omega}\;{\hat \Delta}_j\;
{\bar {\cal G}}_{j\omega},\\
&{\bar {\cal G}}_{j\omega}= {\bar {\cal G}}^{(0)}_{j\omega} + {\bar
{\cal G}}^{(0)}_{j\omega}\;{\hat \Delta}^{\ast}_j \; {\cal
G}^{(0)}_{j\omega}\;{\hat \Delta}_j \;{\bar {\cal G}}_{j \omega},
\end{align}
\end{subequations}
which makes it possible to expand ${\cal F}_{j,\omega}$ in powers of
$\Delta_j$, when working near $T_c$. This is the well-known basis
for Gor'kov's derivation of the GL theory~\cite{gor}.

Using the definition of the anomalous (Gor'kov) Green's function
$$
\frac{1}{\beta\hbar}\sum\limits_{\omega} e^{-i\omega
(t-t')}\langle {\bf x}|{\cal F}_{j\omega}|{\bf x}'\rangle=
-\frac{1}{\hbar}\langle {\cal T}{\hat \psi}_{j\uparrow}({\bf
x}t){\hat \psi}_{j\downarrow} ({\bf x}'t')\rangle,
$$
one can rewrite Eq.~(\ref{gap2band}) in the form
\begin{subequations}\label{basis}
\begin{align}
&\Delta_1({\bf x}) =\lambda_{11} n_1 R_1({\bf x})
+\lambda_{12} n_2 R_2({\bf x}), \label{basisA}\\
&\Delta_2({\bf x}) =\lambda_{21} n_1 R_1({\bf x}) +\lambda_{22} n_2
R_2({\bf x}), \label{basisB}
\end{align}
\end{subequations}
where $R_j({\bf x})$ is a polynomial of $\Delta_j({\bf x})$ and its
spatial derivatives; $\lambda_{ij} = g_{ij}N(0)$ and $n_j =
N_j(0)/N(0)$, where $N_j(0)$ is the band-dependent density of
states, and $N(0)=\sum_j N_j(0)$. To construct the GL equations for
a two-band superconductor, one should evaluate $R_i$ with accuracy
${\cal O}(\tau^{3/2})$. This results in two equations for
$\Delta_1({\bf x})$ and $\Delta_2({\bf x})$ coupled through the
Josephson-like terms (for a clean two-band s-wave superconducting
system, see, e.g., Refs.~\onlinecite{zhit,kog}; for a dirty two-band
superconductor, see, e.g., Refs.~\onlinecite{gur,kosh}). This is
where the aforementioned analysis of Kogan and Schmalian \cite{kog}
is important, as such a representation of the two-band GL equations
must be corrected in order to avoid the appearance of terms of
orders higher than $\tau^{1/2}$ in $\Delta_j$. Appropriate correcting
procedure is given in detail in Ref.~\onlinecite{kog}, and results in
two decoupled GL equations for $\Delta_1$ and $\Delta_2$ which {\it
exactly map on the one-band GL theory}: $\Delta_1({\bf
x})\propto\Delta_2({\bf x})$ and both have the same coherence length
unlike the expectations based on the initial formulation of the
two-band GL formalism.

We now extend the GL formalism up to the order $\tau^{3/2}$ in
$\Delta_j$'s, by taking
\begin{equation}
\Delta_j({\bf x})=\Delta^{(0)}_j({\bf x})+ \Delta^{(1)}_j ({\bf x}),
\label{two-orders}
\end{equation}
with $\Delta^{(0)}_j \propto \tau^{1/2}$ and $\Delta^{(1)}_j \propto
\tau^{3/2}$. To begin with, we limit ourselves to a case of the
zero-magnetic field ($\Delta_j$'s are real). Evaluating $R_j$ with
accuracy ${\cal O}(\tau^{5/2})$, we obtain
\begin{multline}
R_j = -{\widetilde a} \Delta_j - {\widetilde b} \Delta^3_j +
{\widetilde c}\Delta^5_j \\+{\widetilde {\cal K}}_j \nabla^2
\Delta_j + {\widetilde {\cal Q}}_j \nabla^2(\nabla^2 \Delta_j)
-{\widetilde {\cal L}}_j \Delta_i\nabla\cdot (\Delta_j\, \nabla
\Delta_j), \label{Ri}
\end{multline}
with
\begin{align}
&{\widetilde a}=-\bigl({\cal A}+ \tau +
\frac{\tau^2}{2}\bigr),\;{\cal A}=\ln\bigl(\frac{2e^{\Gamma}
\hbar\omega_D}{\pi T_c}\bigr),
\nonumber\\
&{\widetilde b} = W^2_3 (1+2\tau),\;
W^2_3=\frac{7\zeta(3)}{8\pi^2T_c^2}\;\;(W_3 \sim \frac{1}{\pi T_c}),
\nonumber\\
&{\widetilde c}=W^4_5,\; W_5^4=\frac{93\zeta(5)}{128\pi^4T_c^4}
\;\;(W_5 \sim \frac{1}{\pi T_c}),
\nonumber\\
&{\widetilde {\cal K}}_j=\frac{W^2_3}{6}\hbar^2 v^2_j (1+2\tau),
\nonumber\\
&{\widetilde {\cal Q}}_j=\frac{W^4_5}{30}\hbar^4 v^4_j,\;
{\widetilde {\cal L}}_j=\frac{5}{9}W^4_5\hbar^2 v^2_j,
\label{factors}
\end{align}
where $\hbar\omega_D$ is the Debye energy, $\zeta(\ldots)$ is the
Riemann zeta-function, $\Gamma = 0.577$ is the Euler constant, and
the band-dependent Fermi velocity is denoted by $v_j$. Note that, as
compared to the results of Refs.~\onlinecite{zhit} and \onlinecite{kog},
there are three new terms in Eq.~(\ref{Ri}): $\propto \Delta^5_j,
\propto \nabla^2(\nabla^2 \Delta_j)$, and $\propto \Delta_j\nabla
\cdot(\Delta_j\, \nabla\Delta_j)$. In addition, the coefficients
${\widetilde a}, {\widetilde b}$ and ${\widetilde {\cal K}}_j$
contain extra contributions, i.e., ${\widetilde a}$ is now accurate
up to the order $\tau^2$ whereas ${\widetilde b}$ and ${\widetilde{
\cal K}}_j$ include terms $\propto \tau$. Note also that when
evaluating $\Delta_j$'s with accuracy ${\cal O}(\tau^{3/2})$~[see
Eq.~(\ref{two-orders})], we have $\nabla^2 \propto \tau$ in both
$\Delta^{(0)}_j$ and $\Delta^{(1)}_j$.

Going back to Eq.~(\ref{basis}), one obtains $R_1=\Delta_1 -
\lambda_{12} n_2 R_2/(\lambda_{11} n_1)$ from Eq.~(\ref{basisA}),
which can then be inserted into Eq.~(\ref{basisB}). Similarly, $R_2$
can be expressed as $R_2=\Delta_2 -\lambda_{21} n_1
R_1/(\lambda_{22} n_2)$ from Eq.~(\ref{basisB}) and substituted in
Eq.~(\ref{basisA}). Such a manipulation, combined with
Eq.~(\ref{Ri}), results in the following equations
\begin{subequations}
\begin{align}
a_1 \Delta_1 + b_1 \Delta^3_1 - c_1 \Delta_1^5 -{\cal K}_1\nabla^2
\Delta_1^2 - {\cal Q}_1
\nabla^2(\nabla^2\Delta_1)&\nonumber\\
+{\cal L}_1\Delta_1\,\nabla \cdot (\Delta_1\,
\nabla\Delta_1)-\gamma \Delta_2=0,\;& \label{basis1A}\\
a_2 \Delta_2 + b_2 \Delta^3_2 - c_2 \Delta_2^5 -{\cal K}_2\nabla^2
\Delta_2^2 - {\cal Q}_2
\nabla^2(\nabla^2\Delta_2)&\nonumber\\
+{\cal L}_2\Delta_2\,\nabla \cdot (\Delta_2\, \nabla\Delta_2)-\gamma
\Delta_1=0,\;& \label{basis1B}
\end{align}
\end{subequations}
where
\begin{equation}
a_j=\frac{N(0)}{\eta}\Bigl[{\cal A}_j - \eta n_j \bigl(\tau
+\frac{\tau^2}{2}\bigr)\Bigr],\; \eta = \lambda_{11}\lambda_{22} -
\lambda^2_{12}, \label{ai}
\end{equation}
with ${\cal A}_1 = \lambda_{22} - \eta n_1 {\cal A}$ and ${\cal A}_2
= \lambda_{11} - \eta n_2 {\cal A}$~($\eta$ denotes the determinant
of the $\lambda_{ij}$ matrix, where $\lambda_{12}= \lambda_{21}$).
In addition, $b_j, \,c_j,\,{\cal K}_j,\,{\cal Q}_j,\,{\cal L}_j$ in
Eqs.~(\ref{basis1A}) and (\ref{basis1B}) are ${\widetilde
b},\;{\widetilde c},\;{\widetilde {\cal K}}_j,\; {\widetilde {\cal
Q}}_j,\;{\widetilde {\cal L}}_j$ multiplied by $n_jN(0)$,
respectively. The last terms in the left-hand side of
Eqs.~(\ref{basis1A}) and (\ref{basis1B}) are the Josephson interband
coupling terms with $\gamma = \lambda_{12} N(0)/\eta$.

Proceeding in the manner similar to that of Ref.~\onlinecite{kog}, we
now group the terms of the same order in Eqs.~(\ref{basis1A}) and
(\ref{basis1B}). Keeping only terms of the order $\tau^{1/2}$ in
both equations we find
\begin{equation}
\left(\frac{a_1a_2}{\gamma}-\gamma\right)_{\tau^0} =0, \label{tau12}
\end{equation}
where $(\mathcal{B})_{\tau^k}$ denotes the term in the expression
$\mathcal{B}$ of the order $\tau^k$, with $k$ an integer. Equation
(\ref{tau12}) allows one to evaluate $T_c$ in the two-band
superconducting system and is reduced to ${\cal A}_1{\cal A}_2 =
\lambda^2_{12}$, which recovers Eq. (17) from Ref.~\onlinecite{kog}.

Further, when collecting the terms proportional to $\tau^{3/2}$ in
Eqs.~(\ref{basis1A}) and (\ref{basis1B}) we find
\begin{equation}
\alpha \Delta_j^{(0)} + \beta_j [\Delta_j^{(0)}]^3 -
K\nabla^2\Delta_j^{(0)} = 0, \label{tau32}
\end{equation}
where
\begin{align}
&\alpha =\left(\frac{a_1a_2}{\gamma}- \gamma\right)_{\tau},\quad
K=\left(\frac{{\cal K}_1a_2+{\cal K}_2
a_1}{\gamma}\right)_{\tau^0},\nonumber\\
&\beta_1= \left(\frac{b_1a_2+a_1^3b_2/
\gamma^2}{\gamma}\right)_{\tau^0},\quad \beta_2 =
\beta_1\bigl|_{1\leftrightarrow2}, \label{tau32A}
\end{align}
where $\beta_2$ is obtained from the expression for $\beta_1$ by
replacing indices of $a_j$'s and $b_j$'s ($1\leftrightarrow2$).
Equation (\ref{tau32}) is the correct formulation of the standard GL
approach for the two-band $s$-wave clean superconducting system, as
found in Ref.~\onlinecite{kog}. Using Eq.~(\ref{tau12}), we indeed obtain
from Eqs.~(\ref{tau32}) and (\ref{tau32A}) that
\begin{equation}
[\Delta^{(0)}_1({\bf x})/\Delta^{(0)}_2({\bf x})]^2= {\cal
A}_2/{\cal A}_1, \label{prop}
\end{equation}
which follows from the scaling $\beta_1/\beta_2={\cal A}_1/{\cal
A}_2$.

Now, taking the terms of order $\tau^{5/2}$ in Eqs.~(\ref{basis1A})
and (\ref{basis1B}), we arrive at
\begin{multline}
\Delta^{(1)}_j \bigl(\alpha + 3\beta_j
[\Delta_j^{(0)}]^2\bigr)- K\nabla^2\Delta_j^{(1)}\\
=F(\Delta^{(0)}_j) + F_j(\Delta^{(0)}_j), \label{tau52}
\end{multline}
with
\begin{equation}
F(\varphi)=\sigma \varphi + S \nabla^2\varphi + Y
\nabla^2(\nabla^2\varphi), \label{tau52A}
\end{equation}
and
\begin{multline}
F_j(\varphi)=\rho_j \varphi^3 + \chi_j \varphi^5 +U_j\varphi \nabla
\cdot (\varphi \nabla\varphi)
\\+V_j\nabla^2\varphi^3 +Z_j\varphi^2\nabla^2\varphi.
\label{tau52B}
\end{multline}
Equation~(\ref{tau52}) is the first main result of this paper. It
includes all contributions to order $\tau^{3/2}$ to $\Delta_j$'s.
Coefficients $\sigma,\,S$ and $Y$ in Eq.~(\ref{tau52A}) are given by
\begin{align}
&\sigma =-\left(\frac{a_1a_2}{\gamma}- \gamma\right)_{\tau^2},\quad
S=\left(\frac{{\cal K}_1a_2+{\cal K}_2
a_1}{\gamma}\right)_{\tau},\nonumber\\
&Y= \left(\frac{{\cal Q}_1a_2+{\cal Q}_2 a_1-{\cal K}_1{\cal
K}_2}{\gamma} \right)_{\tau^0}, \label{tau52AA}
\end{align}
while the coefficients in Eq.~(\ref{tau52B}) read
\begin{align}
&\rho_1 =-\left(\frac{b_1a_2 + a_1^3b_2/
\gamma^2}{\gamma}\right)_{\tau},\nonumber\\
&\chi_1=\left(\frac{c_1a_2-3a_1^2b_1b_2/\gamma^2 +
a^5_1c_2/\gamma^4}{\gamma}\right)_{\tau^0},
\nonumber\\
&U_1= -\left(\frac{{\cal L}_1a_2+a^3_1{\cal L}_2/
\gamma^2}{\gamma}\right)_{\tau^0},\nonumber\\
&V_1= \left(\frac{b_1{\cal K}_2}{\gamma} \right)_{\tau^0},
Z_1=3\left(\frac{a^2_1{\cal K}_1b_2}{\gamma^3}\right)_{\tau^0},
\label{tau52BB}
\end{align}
and $\rho_2,\,\chi_2,\,U_2$, and $V_2$ are obtained from
Eq.~(\ref{tau52BB}) by replacing $1\leftrightarrow2$ in all relevant
indices.

Now, if the terms $F_j(\Delta^{(0)}_j)$ were absent in
Eq.~(\ref{tau52}), we would obtain that $\Delta^{(1)}_1({\bf x})$ is
proportional to $\Delta^{(1)}_2({\bf x})$ and, furthermore, the
ratio $\Delta^{(1)}_1({\bf x})/\Delta^{(1)}_2 ({\bf x})$ would be
identical to $\Delta^{(0)}_1({\bf x})/\Delta^{(0)}_2 ({\bf x})$
given by Eq.~(\ref{prop}). However, in the presence of
$F_j(\Delta^{(0)}_j)$, this is no longer the case, as not all terms
appearing in $F_j(\Delta^{(0)}_j)$ support the above scaling of the
order parameters. In particular, let us consider the term
$\rho_j[\Delta^{(0)}_j]^3$. This term could support the scaling only
if the ratio $\rho_1/\rho_2$ is equal to ${\cal A}_1/{\cal A}_2$.
From Eq.~(\ref{tau52BB}) we find
\begin{equation}\label{rho}
\frac{\rho_1}{\rho_2}=\frac{{\cal A}_1}{{\cal A}_2}\,\;\frac{2(n_1{\cal
A}^2_2+n_2{\cal A}^2_1)-\eta n_1n_2({\cal A}_2+3{\cal
A}_1)}{2(n_1{\cal A}^2_2+n_2{\cal A}^2_1)-\eta n_1n_2({\cal
A}_1+3{\cal A}_2)},
\end{equation}
which means that $\rho_1/\rho_2 \not={\cal A}_1/{\cal A}_2$ and,
consequently,
\begin{equation}
[\Delta^{(1)}_1({\bf x})/\Delta^{(1)}_2({\bf x})]^2\not= {\cal A}_2
/{\cal A}_1. \label{nonprop}
\end{equation}
Moreover, as seen from the structure of Eq.~(\ref{tau52}), it is
clear that $\Delta^{(1)}_1({\bf x})$ is not at all proportional to
$\Delta^{(1)}_2 ({\bf x})$. We hereby arrive at our main conclusion,
i.e., the band order parameters $\Delta_1({\bf x})$ and
$\Delta_2({\bf x})$ are {\it not proportional} to each other when
extending the Ginzburg-Landau formalism to terms in $\Delta_j$'s
proportional to $\tau^{3/2}$ (beyond the standard terms $\propto
\tau^{1/2}$). This means that the band coherence lengths are {\it in
general different}, and this difference disappears only in the limit
$T\rightarrow T_c$.

For completeness, we give here several remarks about a
generalization of the extended two-band GL formalism to the case of
a nonzero magnetic field (inclusion of a magnetic field will not
affect any of the above conclusions). Such a generalization is not
straightforward because in the first step one needs to go beyond the
eikonal approximation adopted by Gor'kov for the normal state
Green's function~(see, e.g., the textbook~\cite{fetter}). This task
assumes extensive calculations with numerous details that are not
suitable for a Letter. Therefore, we include here only the final
result, while preserving the full derivation for a separate
publication:
\begin{multline}
\langle {\bf x}|{\widetilde {\cal G}}^{\,(0)}_{j\omega}|
{\bf x}'\rangle =
 e^{\frac{i e}{\hbar c}
\int\limits_{{\bf x}^\prime}^{{\bf x}}{\bf A }({\bf r}) d{\bf r}}
\;\biggl\{1+\frac{e^2}{24m^2c^2}\,{\bf B}^2({\bf x})\\
\times\Bigl[\frac{\partial^2}{\partial\omega^2} + \frac{i}{\hbar}
m ({\bf x}- {\bf x}')_{\perp}^2\frac{\partial}{\partial\omega}\Bigr]
\biggr\}\,\langle {\bf x}|{\cal G}^{(0)}_{j\omega}| {\bf x}'\rangle,
\label{eik_ext}
\end{multline}
where ${\widetilde {\cal G}}^{\,(0)}_{j\omega}$ is the normal state
Green's function in the presence of a magnetic field; the integration
in the exponent is taken along a straight line connecting ${\bf x}$
and ${\bf x}'$; $({\bf x}-{\bf x'})_{\perp}$ is the component of the
vector perpendicular to ${\bf B}({\bf x})={\rm rot}{\bf A}({\bf
x})$. As follows from Eq.~(\ref{eik_ext}), the corrections to the
Gor'kov approximation are gauge invariant and of order $\tau^2$~(${\bf
A}\propto\tau^{1/2}$ and ${\bf B} \propto \tau$). In particular,
Eq.~(\ref{Ri}) in the presence of a magnetic field reads
\begin{multline}
R_j = \Bigl[-{\widetilde a} + \frac{W^2_3}{3}\hbar^2\Omega^2({\bf x})\Bigr]
\Delta_j - {\widetilde b}\,\Delta_j |\Delta_j|^2 +{\widetilde c}\,\Delta_j|\Delta_j|^4 \\+{\widetilde {\cal K}}_j {\bf
D}^2\Delta_j + {\widetilde {\cal Q}}_j \Bigl[({\bf D}^2)^2
+\frac{4m^2\Omega^2({\bf x})}{\hbar^2} \\ +\frac{4ie}{3\hbar c}
\;{\rm rot}{\bf B}({\bf x})\,{\bf D}\Bigr]\Delta_j-\frac{{\widetilde {\cal L}}_j}{5}\Bigl[4|\Delta_j|^2{\bf D}^2\Delta_j \\+ 3\Delta^{\ast}_j({\bf D}
\Delta_j)^2 + 2\Delta_j|{\bf D}\Delta_j|^2 + \Delta^2_j ({\bf
D}^2 \Delta_j)^{\ast}\Bigr],
\end{multline}
where ${\bf D}=\nabla + 2\pi i {\bf A}({\bf x})/\Phi_0$~($\Phi_0$ is
the superconducting flux quantum) and $\Omega=|e|B({\bf x})/mc$, with
$B({\bf x})=|{\bf B}({\bf x})|$.

As a final note, we state that our approach differs from the theory
of a local superconductor in a slow varying magnetic field, used in
Refs.~\onlinecite{tewordt,werthamer} (the so-called generalized
Ginzburg-Landau-Gor'kov equations). The approach developed in the
latter papers assumes that the gradients of the order parameter are
small but the order parameter itself can be close to its value at
zero temperature. Instead, we extended the two-band Ginzburg-Landau
formalism up to the order $\tau^{3/2}$~(in $\Delta_i$'s). This
requires to accurately select the necessary terms on the basis of
the proper scaling with $\tau$ of $\Delta_i$'s and their spatial
derivatives. The same holds for the magnetic field and its spatial
derivatives, which, contrary to Refs.~\onlinecite{tewordt,werthamer},
requires to go beyond the eikonal approximation of Gor'kov [see
Eq.~(\ref{eik_ext})].

In summary, by developing the extended GL formalism for a two-band
superconductor: (i) we improved the validity of the Ginzburg-Landau
theory at temperatures away from $T_c$; (ii) we showed that the two
position dependent order parameters in a two-band superconductor are
generally not proportional to each other, thus their spatial scales
are decoupled - contrary to conclusions of the standard GL
formalism; (iii) we developed a useful tool for further theoretical
studies of two-band superconductivity, which also commands
revisiting many earlier works based on the incomplete formulation of
the two-band GL formalism.

\begin{acknowledgments}
This work was supported by the Flemish Science Foundation (FWO-Vl),
the Belgian Science Policy (IAP) and the ESF-INSTANS network.
Discussions with M. D. Croitoru are gratefully acknowledged.
\end{acknowledgments}


\begin{thebibliography}{99}
\bibitem{GL} V. L. Ginzburg and L. D. Landau, Sov. Phys. JETP
{\bf 20}, 1064 (1950).
\bibitem{aran} I. S. Aranson and L. Kramer, Rev. Mod. Phys.
{\bf 74}, 99 (2002).
\bibitem{canf} P. C. Canfield and G. W. Crabtree, Phys. Today
{\bf 56}, 34 (2003).
\bibitem{pnic} J. Paglione and R. L. Greene, Nat. Phys. {\bf
6}, 645 (2010); M. L. Teague {\it et al.}, arXiv:1007.5086v2.
\bibitem{mosh} V. V. Moshchalkov, M. Menghini, T. Nishio,
{\it et al.}, Phys. Rev. Lett. {\bf 102}, 117001 (2009).
\bibitem{nishio} T. Nishio, V.-H. Dao, Q. H. Chen, {\it et al.},
Phys. Rev. B {\bf 81}, 020506 (2010).
\bibitem{zhit} M. E. Zhitomirsky and V.-H. Dao, Phys. Rev.
B {\bf 69}, 054508 (2004).
\bibitem{gur} A. Gurevich, Phys. Rev. B {\bf 67}, 184515 (2003).
\bibitem{kosh} A. A. Golubov and A. E. Koshelev, Phys. Rev. B
{\bf 68}, 104503 (2003).
\bibitem{babaev} E. Babaev and M. Speight, Phys. Rev. B {\bf 72},
180502 (2005).
\bibitem{babaev1} E. Babaev, J. J\"{a}ykk\"{a}, and M. Speight,
Phys. Rev. Lett. {\bf 103}, 237002 (2009); E. Babaev, J.
Carlstr\"{o}m, and M. Speight, Phys. Rev. Lett. {\bf 105},
067003 (2010).
\bibitem{wang} J.-P. Wang, Phys. Lett. A {\bf 374}, 58 (2009).
\bibitem{brandt} E. H. Brandt and M. P. Das, arXiv:1007.1107v1.
\bibitem{kog} V. G. Kogan and J. Schmalian, arXiv:1008.0581v1.
\bibitem{kopnin} N. B. Kopnin, {\itshape Theory of Nonequilibrium
Superconductivity} (Clarendon Press, Oxford, 2001).
\bibitem{gor} L. P. Gor'kov, Sov. Phys. JETP {\bf 36}, 1364 (1959).
\bibitem{fetter} A. L. Fetter and J. D. Walecka, {\it Quantum
Theory of Many-Particle Systems} (Dover, New York, 2003).
\bibitem{tewordt} L. Tewordt, Phys. Rev. {\bf 132}, 595 (1963).
\bibitem{werthamer} N. R. Werthamer, Phys. Rev. {\bf 132}, 663
(1963).
\end{thebibliography}
\end{document}